\documentclass[twocolumn,apsrev,prb,reprint,superscriptaddress]{revtex4-2}
\usepackage{graphicx,amssymb,amsmath,amsfonts}% Include figure files
\usepackage{tabularx}
\usepackage{bm}% bold math
\usepackage{mathrsfs}
\usepackage{appendix}
\usepackage{bbm}
\usepackage{color}
\usepackage{hyperref}
\usepackage{natbib}

\def \vk {{\bf k}}

\def \vS {{\bf S}}

\def \vh {{\bf h}}
\def \vn {{\bf n}}
\def \vz {{\bf z}}

\def \bsig {{\boldsymbol{\sigma}}}

\usepackage{calligra}
\DeclareMathAlphabet{\mathcalligra}{T1}{calligra}{m}{n}
\DeclareFontShape{T1}{calligra}{m}{n}{<->s*[2.2]callig15}{}
\newcommand{\scripty}[1]{\ensuremath{\mathcalligra{#1}}}

\begin{document}

\title{Topological Magnons on the Ferromagnetic Zigzag Lattice}

\author{Skandan Subramanian}
\email[
This manuscript has been authored in part by UT-Battelle, LLC, under contract DE-AC05-00OR22725 with the US Department of Energy (DOE). The US government retains and the publisher, by accepting the article for publication, acknowledges that the US government retains a nonexclusive, paid-up, irrevocable, worldwide license to publish or reproduce the published form of this manuscript, or allow others to do so, for US government purposes. DOE will provide public access to these results of federally sponsored research in accordance with the DOE Public Access Plan (http://energy.gov/downloads/doe-public-access-plan)]{}
\affiliation{Materials Science and Technology Division, Oak Ridge National Laboratory, Oak Ridge, Tennessee 37831, USA}
\affiliation{Department of Physics, Indian Institute of Technology, Madras, India}
\author{Tom Berlijn}
\affiliation{Center for Nanophase Materials Sciences, Oak Ridge National Laboratory, Oak Ridge, Tennessee 37831, USA}
\author{Lucas Lindsay}
\affiliation{Materials Science and Technology Division, Oak Ridge National Laboratory, Oak Ridge, Tennessee 37831, USA}
\author{Randy S. Fishman}
%\email[r.fishman@unf.edu]{}
\affiliation{Materials Science and Technology Division, Oak Ridge National Laboratory, Oak Ridge, Tennessee 37831, USA}
\author{John W. Villanova}
\email[john.villanova@mtsu.edu]{}
\affiliation{Department of Physics and Astronomy, Middle Tennessee State University, Murfreesboro, Tennessee 37132, USA}

\date{\today}

\begin{abstract}  
Motivated by the experimental identification of magnetic compounds consisting of zigzag chains, we analyze the band structure topology of magnons in ferromagnets on a zigzag lattice. We account for the general lattice geometry by including spatially anisotropic Heisenberg exchange interactions and by Dzyaloshinskii-Moriya interaction on inversion asymmetric bonds. Within the linear spin-wave theory, we find two magnon branches, whose band structure topology (i.e., Chern numbers) we map out in a comprehensive phase diagram. Notably, besides topologically trivial and gapless phases, we identify topologically nontrivial phases that support chiral edge magnons. We show that these edge states are robust against elastic defect scattering.
\end{abstract}
%We study the topological behavior of a ferromagnetic zigzag model with exchange interactions $J_{nx}$ and $J_{ny}$ along adjacent $n=1$ and 2 chains. Due to the different exchange interactions on adjacent chains, this model supports Dzyaloshinskii-Moriya interactions $D$ along the $(1,-1)$ diagonals.   In an isotropic model with $J_{1x}=J_{2x}$ and $J_{1y}=J_{2y}$, the magnon bands are degenerate along the boundaries of the first Brillouin zone with $k_x-k_y =\pm \pi/a$ and the Chern numbers $C_n$ are not well defined.  However, an anisotropic model with $J_{1y}\ne J_{1x}$ lifts that degeneracy and produces well-defined Chern numbers of $C_n=0$ or $\pm 1$ for the two well-separated magnon bands. By mapping our model onto a Haldane model we show that the Chern numbers only depend on the parameters $\alpha =\vert (J_{1y}-J_{2y})/(J_{1x}-J_{2x})\vert $ and $\beta = \vert (J_{1y}+J_{2y})/(J_{1x}+J_{2x})\vert $. We also show the edge states of the ferromagnetic zigzag lattice cut onto a ribbon are protected from disorder.


\maketitle

\section{Introduction}
Topology has become a critical topic in condensed matter physics \cite{HasanRev2010,QiRev2011,McClartyRev2022}. Toplogical edge states have drawn significant attention due to their potential for enabling dissipationless and robust channels for information transfer \cite{Pantaleon2018,Mei2019,Malz2019,Zhang2020}. Recently the concept of topological band structures has been extended to the elementary excitations of magnetically ordered insulators, magnons. Magnonic topological insulators, characterized by nontrivial topological invariants, have been predicted \cite{Fujimoto2009,Cao2015,Zhu2021,McClartyRev2022}. A major experimental challenge lies in directly detecting topological edge states, as neutron scattering cannot capture their signature. Thus local spectroscopic probes and transport measurements are important signatures: these topological materials exhibit the magnon Hall \cite{Onose2010, Ideue12, Hirschberger15a, Hirschberger15b, Murakami17, Neumann22} and Seebeck \cite{Uchida10, Wu16} effects due to their Berry curvature, which can be considered as a fictitious magnetic field in momentum space \cite{Chang96,Sundaram99,Xiao10}. 

The identification of suitable material candidates remains a critical bottleneck. Honeycomb lattice van der Waals magnets represent some of the most prominent current examples of potential magnon Chern insulators \cite{McGuire15,Chen18,Joshi2018,Zhu2021}. Unfortunately, many of these materials are highly air-sensitive which complicates experimental studies. Expanding the class of materials capable of hosting magnon Chern insulators with chiral edge states is therefore an urgent priority for advancing the field. Previous studies have demonstrated that magnon Chern insulators can be realized in ferromagnets on various lattice geometries, including the pyrochlore \cite{Onose2010, Ideue12}, kagome \cite{Katsura2010}, star \cite{Owerre17}, and honeycomb \cite{Cheng16, Owerre16, Fishman23b} lattices. Recently, the zigzag lattice has also been identified as a platform with promising topological properties of magnons \cite{Fishman23c}. In this work, we conduct a detailed analysis of the topological properties of magnons in zigzag lattice ferromagnets. We derive analytic conditions for the exchange interactions that enable topologically nontrivial magnon phases. Furthermore, we investigate the robustness of chiral edge states against disorder. Our findings contribute to broadening the set of candidate materials for realizing magnon Chern insulators, thereby paving the way for future experimental and theoretical exploration.

\begin{figure}
\begin{center}
\includegraphics[width=5cm]{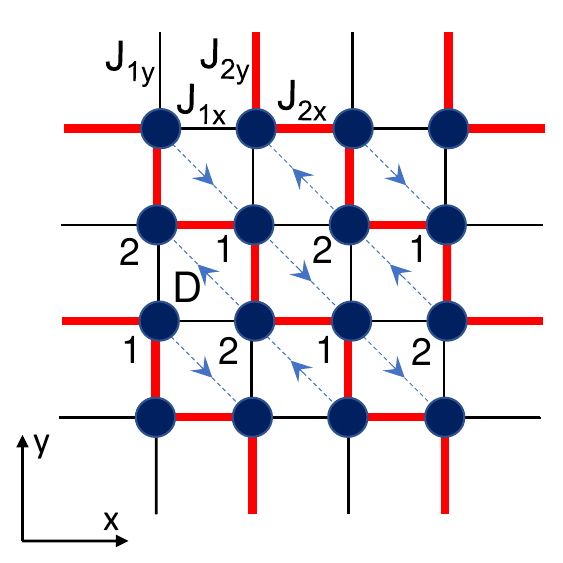}
\end{center}
\caption{A FM zigzag lattice with four different exchange parameters.  The DM interaction $-D(\vS_i\times \vS_j)\cdot \vz $ lies along the $(1,-1)$ diagonals with the arrows
in the figure pointing from $\vS_i$ to $\vS_j$.}
\label{Fig1}
\end{figure}

The structure of the ferromagnetic (FM) zigzag lattice is sketched in Fig.~\ref{Fig1}, which has alternating black and red bonds with exchange couplings $J_{1\alpha }$ and $J_{2\alpha }$, respectively, where the $\alpha \in \{x,y\}$ exchanges can be different.
We define
\begin{equation}
r=\frac{J_{2x}+J_{2y}}{J_{1x}+J_{1y}},
\label{AR}
\end{equation}
so that the directionally-averaged $J_2>0$ is larger than the directionally-averaged $J_{1}>0$ when $r>1$.
Several candidate FM zigzag materials have been identified: spin-1/2 Heisenberg vanadium chains in CdVO$_3$ \cite{Onoda99, Tsirlin11}, spin-3/2 chromium chains in LaCrOS$_2$ \cite{Takano02}, and spin-3.4/2 manganese chains in La$_3$MnAs$_5$ \cite{Duan22}. In all of these materials $r>1$. For CdVO$_3$ \cite{Onoda99, Tsirlin11}, $J_2\approx 90$\,K and $J_1 \approx 18$\,K so that $r\approx 5$.   For La$_3$MnAs$_5$ \cite{Duan22}, $r \approx 7.6$. Similarly, $r$ is also believed to be large in LaCrOS$_2$ \cite{Takano02}. However, the magnitude of the anisotropy $\xi_1=J_{1x}/J_{1y}$ and $\xi_2=J_{2x}/J_{2y}$ in these materials is unknown.

As we shall see, the topological properties in this lattice are controlled by the relative sizes of the exchange parameters. These properties include the size of the gap (if any) between the magnonic bands and the Chern numbers of the bands (either $C_n=\pm 1$ or $C_n=0$). Associated with the Chern numbers are the number of edge modes on a ribbon constructed from the material \cite{Mat11a, Mat11b, Mong11, Zhang13, Mook14a}. Strain can be used to conveniently control the anisotropy of the exchange parameters and the topological properties of FM zigzag lattice materials.

\section{Model and Method}

Because $J_{1\alpha }\ne J_{2\alpha }$, inversion symmetry is broken across the diagonals $[1,-1]$ of Fig.~\ref{Fig1}.   
Traversing such a diagonal, the environment to the upper right (consisting of $J_{2\alpha }$ bonds) is different from the environment to the lower left 
(consisting of $J_{1\alpha }$ bonds).  Due to this broken inversion symmetry, Dzyaloshinskii-Moriya (DM) interactions can couple the spins on either end of the diagonal dashed lines.  We denote this interaction by $D$ and the interaction by $-D(\vS_i \times \vS_j)\cdot \vz $.  
%For the lines coupling sites 1, sites $j$ are to the lower right of sites $i$;  for the lines coupling sites 2, sites $j$ are to the lower left of sites $i$.  These directions are indicated by the arrows in Fig.\,1.

Broken inversion symmetry also allows another DM interaction $D'$ along $[1,1]$ between neighboring red chains, but this interaction is generally weaker than 
$D$ (because it involves inversion symmetry broken over larger distances in real materials) and only serves to complicate the model.

The Hamiltonian of the FM zigzag lattice is then given by
\begin{equation}
\begin{aligned}
{\cal H}=&-\frac{1}{2} \sum_{i,j} J_{ij}\,\vS_i \cdot \vS_j -D\sum_{i,j}(\vS_i\times \vS_j)\cdot \vz \\
&- K\sum_i (\vS_i \cdot \vz )^2,
\end{aligned}
\end{equation}
where $K>0$ is a single-ion anisotropy that tends to align the spins along the $z$-axis. Notice that there are two different magnetic sites in each unit cell. The DM interaction $D$ couples the spin on one type of magnetic site to the spin on the same type of magnetic site in neighboring unit cells (1 to 1 or 2 to 2), whereas the exchange interactions, $J_{ij}$, couple nearest-neighbor sites of two different types (1 to 2).

We map the spin operators to Boson operators by means of the Holstein-Primakoff transformation: $S_{iz}=S-a_i^{\dagger }a_i$, $S_{i+}=S_{ix}+iS_{iy}=\sqrt{2S}a_i$, and $S_{i-}=S_{ix}-iS_{iy}=\sqrt{2S}a_i^{\dagger }$. The Hamiltonian is rewritten in this new form as
\begin{equation}
{\cal H} =\sum_{\vk } {\underline v}(\vk )^{\dagger } \cdot \underline{H}(\vk ) \cdot {\underline v}(\vk ),
\label{HV}
\end{equation}
\begin{equation}
{\underline v}(\vk ) =\left( \begin{array}{c} a_{\vk }^{(1)} \\ a_{\vk }^{(2)} 
\end{array} \right),
\end{equation}
\begin{equation}
{\underline H}(\vk )= \frac{J_t S}{2}  \left( \begin{array}{cc} A_{\vk}^- & -\Psi_{\vk }^* \\
-\Psi_{\vk }& A_{\vk^+}  
\end{array} \right) ,
\label{HW}
\end{equation}
where
\begin{equation}
A_{\vk }^{\pm }=1\pm d\tau_{\vk } +\kappa ,
\end{equation}
\begin{equation}
J_t=J_{1x}+J_{1y}+J_{2x}+J_{2y},
\end{equation}
\begin{equation}
\Psi_{\vk }=\frac{J_{1x}+J_{1y}e^{i(k_x-k_y)a}+J_{2x}e^{2ik_xa}+J_{2y}e^{i(k_x+k_y)a}}{J_t},
\end{equation}
$d=-4D/J_t$, $\kappa =2K/J_t$,
and $\tau_{\vk } = \sin ((k_y-k_x)a)$.

As found earlier \cite{Fishman23c}, the excitation frequencies are
\begin{equation}
\hbar \omega_1(\vk)=J_tS ( 1  - \mu_{\vk } + \kappa )
\label{zzfm1r},
\end{equation}
 \begin{equation}
\hbar \omega_2(\vk)=J_tS ( 1  + \mu_{\vk } + \kappa ),
\label{zzfm2r}
\end{equation}
where
\begin{equation}
\mu_{\vk }=\sqrt{\vert \Psi_{\vk }\vert^2 +(d\, \tau_{\vk } )^2}.
\end{equation}
Notice that the only role of the anisotropy is to shift the excitation frequencies upwards by $2KS$.

The first Brillouin zone (BZ) of the FM zigzag lattice is a diamond constructed by the lines $k_x-k_y =\pm \pi/a$ and $k_x+k_y =\pm \pi/a$.  For the isotropic model with $\xi_1=\xi_2=1$, we find that $\tau_{\vk }=0$, $\Psi_{\vk }=0$, $\mu_{\vk }=0$, and $\omega_1(\vk )=\omega_2(\vk)$ at the boundary $k_x-k_y =\pm \pi/a$.
The system is gapless. When the model is anisotropic with $\xi_1\ne1 $ or $\xi_2\ne 1$, then $\tau_{\vk }$ still vanishes at this
boundary of the first BZ but $\Psi_{\vk }$ and $\mu_{\vk }$ do not, and the modes are non-degenerate except possibly at isolated $\vk $ points along $k_x-k_y =\pm \pi/a$. The system is generically gapped and insulating except possibly at isolated $\vk $ points for certain values of the exchange parameters.

We comment at this stage that the Hamiltonian we have defined here differs slightly from that in Ref.~\cite{Fishman23c} in the phase factors appearing in $\Psi_{\vk}$. This amounts to the difference, familiar to electronic tight-binding models, between phase factors relating hopping between unit cells, $e^{-i\vk\cdot{\bf R}}$, versus including information about the specific site,\,$\scripty{r}_j$, within a unit cell, $e^{-i\vk\cdot({\bf R}+{\bf \scripty{r}_j})}$. The former is widely used since there is no effect on the eigenvalues of the problem and one need only retain a record of which unit cell is the endpoint for the hopping (e.g. Wannier90 \cite{w90}). However, the site-discerning phase factor is sometimes favored (e.g. PythTB \cite{pythtb}), and defining the Hamiltonian this way ensures that the surface mapped by $\vh(\vk)$ (defined next) is closed upon scanning the first BZ only once.

We now derive a condition for the bulk system to be a Chern insulator with nonzero Chern numbers $C_n$ for the two magnon bands. We write the Hamiltonian matrix in Eq.\,(\ref{HW}) in the Pauli matrix basis
\begin{equation}
\underline{H}(\vk) =\vh(\vk) \cdot {\underline {\bsig }} + h_0 (\vk )\underline{I},
\end{equation}
where $\vh(\vk)$ is a vector, ${\underline {\bsig }}$ the vector of $2\times2$ Pauli matrices, and $\underline{I}$
is the $2\times 2$ identity matrix. The problem is mapped to a form well-suited for analysis of the topology; we follow the geometrical approach that was applied in Ref.~\cite{Fruchart13} to solve the Haldane model \cite{Haldane88}.
Solving for $\vh (\vk)$, we find
\begin{equation}
\begin{aligned}
h_x(\vk ) = -\frac{1}{J_t}\bigg(&J_{1x}+J_{1y}\cos((k_x-k_y)a)\\
&+J_{2x}\cos(2k_xa)+J_{2y}\cos((k_x+k_y)a) \bigg),
\end{aligned}
\end{equation}
\begin{equation}
\begin{aligned}
h_y(\vk ) = -\frac{1}{J_t}\bigg(&J_{1y}\sin((k_x-k_y)a)\\
&+J_{2x}\sin(2k_xa)+J_{2y}\sin((k_x+k_y)a) \bigg),
\end{aligned}
\end{equation}
and
\begin{equation}
h_z(\vk ) =\frac{4D}{J_t}\sin((k_y-k_x)a).
\end{equation}
Our results do not depend on $h_0(\vk )$, which acts like a chemical potential, so long as it is small enough not to close the gap anywhere in the BZ. The overall scale factor of $J_tS/2$ is not relevant to discussion of the topology.

\begin{figure*}
\begin{center}
\makebox[\textwidth]{\includegraphics[width=0.9\textwidth]{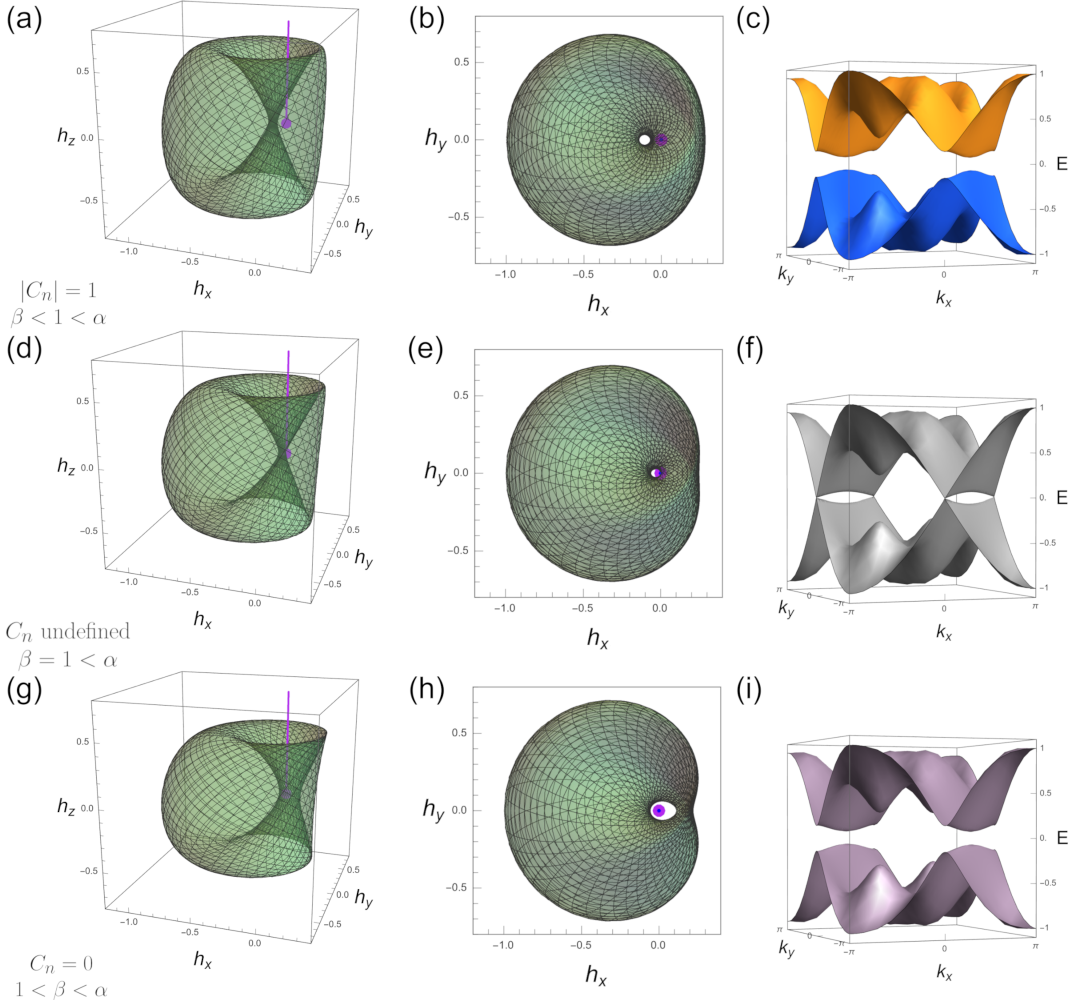}}
\end{center}
\caption{Topology of the Hamiltonian and band structure. (a) Side and (b) top views of the $\vh$-surface for the model with $\{\alpha,\beta\}=\{\frac{5}{3},\frac{3}{4}\}$. A +$h_z$-ray intersects the surface once, corresponding to $|C_1|=1$. (c) The band structure exhibiting a nontrivial gap; the color of the bands emphasizes the opposite Chern number of the bands. (d-f) The same panels, respectively, for the model with $\{\alpha,\beta\}=\{\frac{5}{3},1\}$ and undefined Chern number. The color of the (gapless) bands emphasizes the undefined Chern number. (g-i) The same panels, respectively, for the model with $\{\alpha,\beta\}=\{\frac{5}{3},\frac{4}{3}\}$ and $C_1=C_2=0$.}
\label{Fig2}
\end{figure*}

%We now derive a condition for the bulk system to be a Chern insulator with nonzero Chern numbers $C_n$ for the two magnon bands. We write the Hamiltonian matrix in Eq.\,(\ref{HW}) in the Pauli matrix basis
%\begin{equation}
%\underline{H}(\vk) =\vh(\vk) \cdot {\underline {\bsig }} + h_0 (\vk )\underline{I},
%\end{equation}
%where $\vh(\vk)$ is a vector, $\underline{\sigma}_i$ is a $2\times 2$ Pauli matrix, and $\underline{I}$ is the $2\times 2$ identity matrix. The problem is mapped to a form well-suited for analysis of the topology; this was used for the Haldane model \cite{Haldane88}. Solving for $\vh (\vk)$, we find
%\begin{equation}
%h_x(\vk ) = 
%-\frac{J_{1x} + J_{2x}}{J_t} \cos(k_xa) - \frac{J_{1y} + J_{2y}}{J_t} \cos(k_ya),
%\end{equation}
%\begin{equation}
%h_y(\vk ) =
%\frac{J_{1x} - J_{2x}}{J_t} \sin(k_xa) + \frac{J_{1y} - J_{2y}}{J_t} \sin(k_ya),
%\end{equation}
%and
%\begin{equation}
%h_z(\vk ) =-d\,\tau_{\vk }.
%\end{equation}
%Our results do not depend on $h_0(\vk )$, which acts like a chemical potential. Without loss of generality, we assume that $\underline{H}(\vk )$ is scaled by $2/(SJ_t)$. 

\section{Discussion}
\subsection{Topology of the Bulk Hamiltonian}
The eigenvalues of $\underline{H}(\vk)$ are $\pm \vert \vh(\vk)\vert $, which agrees with the excitation frequencies in  Eqs.\,(\ref{zzfm1r}) and (\ref{zzfm2r}). Adopting a similar methodology to that in Ref.\,\cite{Fruchart13}, we calculate the Chern number of the lower positive frequency band as half the number of intersections of a line through the origin with the $\vh $-surface. A nontrivial Chern number is obtained only if $\vh(\vk)/\vert \vh(\vk)\vert $ completely covers the Bloch sphere $\mathbb{S}^2$ and a smooth function cannot connect every point on $\mathbb{S}^2$ to a specific eigenvector without producing discontinuities at the poles. On the other hand, a trivial topology occurs if the sphere $\mathbb{S}^2$ is not fully covered because we can then define a single map from parameter space to each eigenvector. Either way, the topology is quantified by calculating the Chern number.

The Chern number is given by the expression

\begin{equation}
C =\frac{1}{2}\sum_{\vk \,\epsilon \,\mathbb{D}}
{\rm{sgn}} (h_z(\vk)) \, {\rm{sgn}}\Biggl(
\biggl( \frac{\partial \vh(\vk )}{\partial k_x} \times \frac{\partial \vh (\vk )}{\partial k_y }\biggr)\cdot \vz\Biggr),
\label{C1}
 \end{equation}
where $\mathbb{D}$ is the set of solutions to the equations $h_x(\vk ) =h_y(\vk) = 0$ while demanding $h_z(\vk)\ne 0$. The normal vector $\vn (\vk )$ to the $\vh$-surface is defined as
\begin{equation}
\vn (\vk )=\frac{\partial \vh(\vk )}{\partial k_x} \times \frac{\partial \vh (\vk )}{\partial k_y }.
\end{equation}
Because both partial derivatives $\partial \vh(\vk )/\partial k_x$ and $\partial \vh(\vk )/\partial k_y$ are tangential to the surface, 
the cross product $\vn (\vk )$ is normal to the surface. Implicit in our system of equations is choosing the $h_z$-axis as the line of intersection for the $\vh$-surface. Such intersections with the $h_z$-axis are considered to be positive if $\vn (\vk )$ and $h_z(\vk)$ have the same sign and negative otherwise. Alternatively, one may look only along a +$h_z$-ray and neglect the factor of one-half in Eq.~\ref{C1}. The Chern number is nontrivial if the origin is fully enclosed by the $\vh$-surface (is interior to the surface). Similar to the winding number of a curve on a plane, the result does not depend on the choice of the line mentioned above.

With the definitions
$A=J_{1x}-J_{2x}$, $B=J_{1y}-J_{2y}$, $F=-(J_{1x}+J_{2x})$, and $G=-(J_{1y}+J_{2y})$, we 
solve the system of equations for the condition for $\vk \in \mathbb{D}$:
%\begin{equation}
%\sin (k_xa) = \pm \sqrt{ \frac{1-(G/F)^2}{1-(G/F)^2(A/B)^2} },
%\end{equation}
%\begin{equation}
%\sin(k_ya)= \mp {\rm{sgn}} \biggl( \frac{B}{A} \biggr) \sqrt{ \frac{ 1-(G/F)^2 }{(B/A)^2-(G/F)^2} }.
%\end{equation}
\begin{equation}
k_xa = \text{arctan}\bigg[\mp\frac{B}{G}\sqrt{\frac{G^2-F^2}{A^2-B^2}} \bigg],
\end{equation}
\begin{equation}
k_ya=-\text{arctan}\bigg[\pm\frac{A}{F}\sqrt{\frac{G^2-F^2}{A^2-B^2}} \bigg],
\end{equation}
where the upper sign and lower sign are chosen simultaneously for each ($k_x$, $k_y$) pair. There are four solutions to the system of equations: one is within and one is outside the first BZ, and the other two solutions are ($-k_x$, $-k_y$) in either of these cases. The set $\mathbb{D}$ is composed of only those two solutions inside the first BZ, one for $h_z(\vk)>0$ and one for $h_z(\vk)<0$. If the radicand is not real, then $\mathbb{D}$ is empty and the Chern numbers are 0. To obtain a nontrivial solution for the Chern numbers, we require the conditions
\begin{equation}
\biggl(\frac{G}{F}\biggr)^2 \ge 1 \ge \biggl( \frac{B}{A}\biggr)^2,
\end{equation}
or
\begin{equation}
\biggl(\frac{B}{A}\biggr)^2 \ge 1 \ge \biggl(\frac{G}{F}\biggr)^2.
\end{equation}
%The $z$-component of the normal vector at the point(s) of intersection is:
%\begin{eqnarray}
%n_z(\vk) &=&\frac{a^2}{J_t^2}\Bigl\{ GA\sin(k_ya) \cos(k_xa) \nonumber \\
%&&- BF \sin(k_xa) \cos(k_ya)\Bigr\}.
%\end{eqnarray}
The Chern number is undefined if either $n_z(\vk)=0$ or $h_z(\vk)=0$.
Otherwise the Chern number is nonzero provided that $\mathbb{D}$ is non-empty. Notice that $n_z(\vk)$ goes to zero if either $(B/A)^2 = 1$ or $(G/F)^2 = 1$.
Hence, the natural parameters of the topological phase diagram are $\alpha = \vert B/A \vert $ and $\beta = \vert G/F \vert $.  For the isotropic model, $\alpha =\beta =1$; the magnon bands are gapless and $C$ is undefined.

\begin{figure}
\begin{center}
\includegraphics[width=7cm]{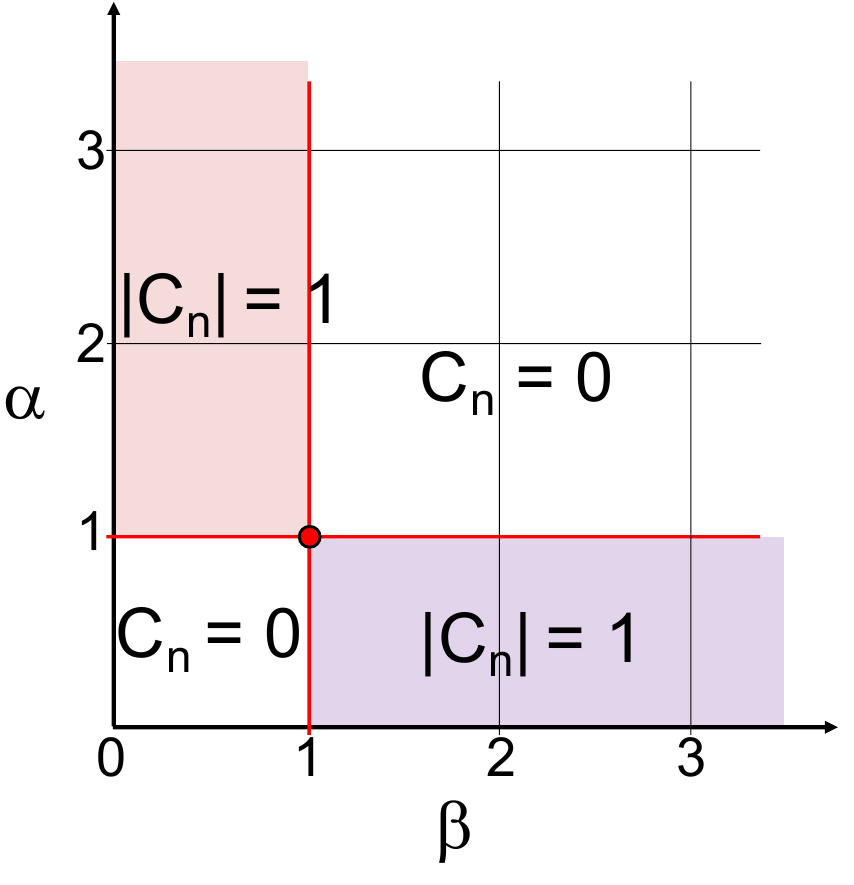}
\end{center}
\caption{The phase diagram of the FM zigzag lattice.  $C_n$ are undefined along the red lines.  
The model is isotropic at the red circle with $\alpha =\beta =1$.}
\label{Fig3}
\end{figure}

Figure~\ref{Fig2}(a)-(c) exhibit a nontrivial band topology. For $\alpha =5/3$ and $\beta =3/4$, the +$h_z$-ray intersects the surface once, as seen in Fig.~\ref{Fig2}(a).
The two bands sketched in Fig.~\ref{Fig2}(c) are gapped and $C_n=\pm 1$.
In Figs.~\ref{Fig2}(d)-(f) with $\alpha =5/3$ and $\beta =1$, the surface generated by $\vh (\vk )$ passes through the origin so the Chern number is undefined. 
The band structure is then gapless.
In Fig.~\ref{Fig2}(g)-(i) with $\alpha = 5/3$ and $\beta  =4/3$, the +$h_z$-ray has no intersections with the surface generated by $\vh (\vk )$ due to the presence of a hole. The bands exhibit a trivial gap; $\mathbb{D}=0$ and $C_n=0$.  

The phase diagram corresponding to these possible cases is shown in Fig.~\ref{Fig3}.  Red lines indicate the cases $\alpha =1$ or $\beta =1$ where the Chern numbers are undefined due to a gapless band structure.
A red dot indicates the point $\alpha =\beta =1$ where the model
is isotropic. For well-defined Chern numbers of $C_n=0$ or $C_n=\pm 1$, the band structure is gapped and the material is insulating. Notice that the phase diagram depends only on the presence of a DM interaction rather than its size, which is not specified in Fig.~\ref{Fig3}. For any 
value of $D$, the retention of the spins along the $z$-axis 
can be guaranteed by the easy-axis anisotropy $K$.

These exact results for the Chern numbers agree with the numerical results obtained in Ref.\,\cite{Fishman23c} based on the Berry curvature, given in a semi-classical description by
\begin{equation}
{\bf \Omega}_n (\vk)=\frac{i}{2 \pi} 
\biggl\{ \frac{\partial }{\partial \vk } \times
\langle u_n(\vk ) \vert \frac{\partial }{\partial \vk }\vert u_n(\vk) \rangle \biggr\}
\label{EqBerry}
\end{equation}
where $\vert u_n(\vk )\rangle $ is the Bloch function.
The Chern number for band $n$ is then given by
\begin{equation}
C_n=\int_{{\rm BZ}} d^2k\, \Omega_{nz}(\vk ).
\end{equation}
It is shown in Ref.\,\cite{Fruchart13} that this reduces to Eq.\,(\ref{C1}) for a two-band model.
Based on the analytic solutions for the excitation frequencies given in 
Eqs.\,(\ref{zzfm1r}) and (\ref{zzfm2r}) and the Berry curvature obtained from spin-wave theory, we numerically evaluated the Chern numbers by performing integrals over the first BZ \cite{Shindou13}. The numerical results agreed with the exact results derived here and presented in the phase diagram of Fig.~\ref{Fig3}.

\begin{figure}
\begin{center}
\includegraphics[width=7cm]{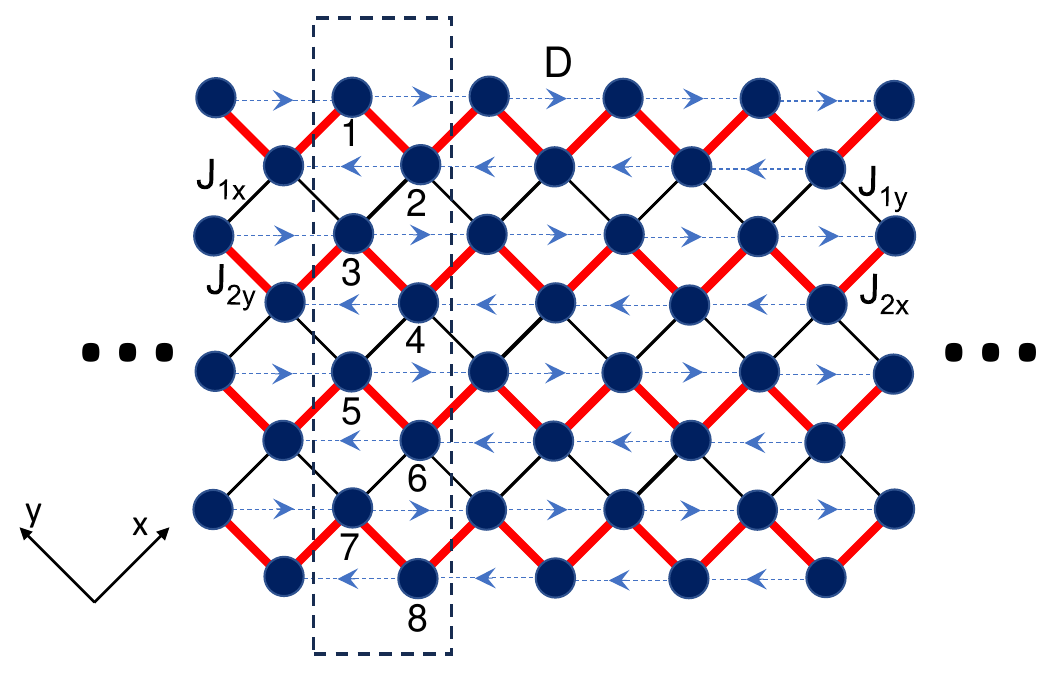}
\end{center}
\caption{A ribbon of zigzag width 8 constructed from the FM zigzag lattice.  The magnetic unit cell contains the 8 sites indicated within the dashed box.}
\label{Fig4}
\end{figure}

\subsection{Edge modes in pristine and defected ribbons}

We now consider the case where a bulk FM zigzag lattice is terminated into a ribbon by making two parallel 45$^\circ$ cuts. Sites lying on the top and bottom edges of the ribbon belong to the same sublattice. An example is sketched in Fig.~\ref{Fig4}, which shows a ribbon with width $M=8$ (8 atomic layers). Notice that the DM interactions lie along the length of the ribbon.  The only valid wavevector for the ribbon is $k_x-k_y$ along its length.

\begin{figure}
\begin{center}
\includegraphics[width=7cm]{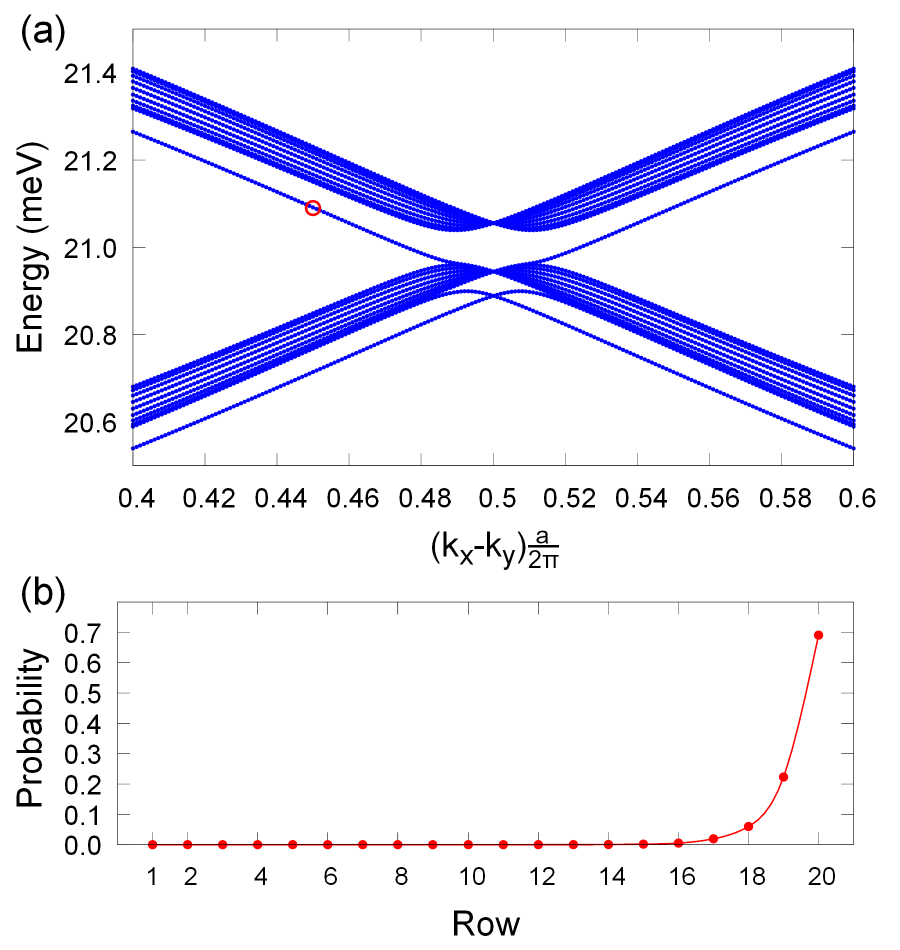}
\end{center}
\caption{(a) Magnon bands for a ribbon of width $W=20$ with $J_{1x}=0.5$, $J_{1y}=1.5$, $J_{2x}=8$, and $J_{2y}=8$. (b) Probability distribution for the layer occupation of the edge state corresponding to the point marked in (a).}
\label{Fig5}
\end{figure}

The excitation frequencies are evaluated by constructing the Hamiltonian matrix $\underline{H}(\vk )$ \cite{fishmanbook18},
which has dimensions $M\times M$ in the absence of defects.  
Because $M \gg 1$, the excitation frequencies must be solved by
numerically diagonalizing the $M\times M$ matrix $\underline{H}(\vk )$.

Due to the bulk-boundary correspondence \cite{Mat11a, Mat11b, Mong11, Zhang13, Mook14a}, the Chern numbers evaluated in the bulk give the number of topological edge modes for the two-dimensional ribbon. 
As expected, systems with undefined $C_n$ are gapless and do not support edge states along the ribbon. Systems with $C_n=0$ exhibit a band gap but also have no edge states. Systems with $C_n=\pm 1$ exhibit a gap between upper and lower bands and also support edge states that connect the upper and lower bands.  

We begin by studying a ribbon without defects.  Taking $J_{1x}=0.5$, $J_{1y}=1.5$, $J_{2x}=8$, and $J_{2y}=8$,
we have $\alpha =\vert B/A\vert <1$ and $\beta=\vert G/F\vert >1$. Based on the phase diagram of Fig.~\ref{Fig3}, these values imply that $C_n=\pm 1$. Hence, the system 
possesses an energy gap and a single state at each edge.  

Those predictions are verified in Fig.~\ref{Fig5}(a), which plots the magnon band energies versus $(k_x-k_y)a/2\pi $ for a ribbon with width $M=20$.  
In Fig.~\ref{Fig5}(b), we plot the layer probability for the edge state that travels around the ribbon with momentum reversed on opposite edges.
This probability map verifies that the edge state resides at one edge with swift decay into the bulk of the ribbon.

\begin{figure}
\begin{center}
\includegraphics[width=7cm]{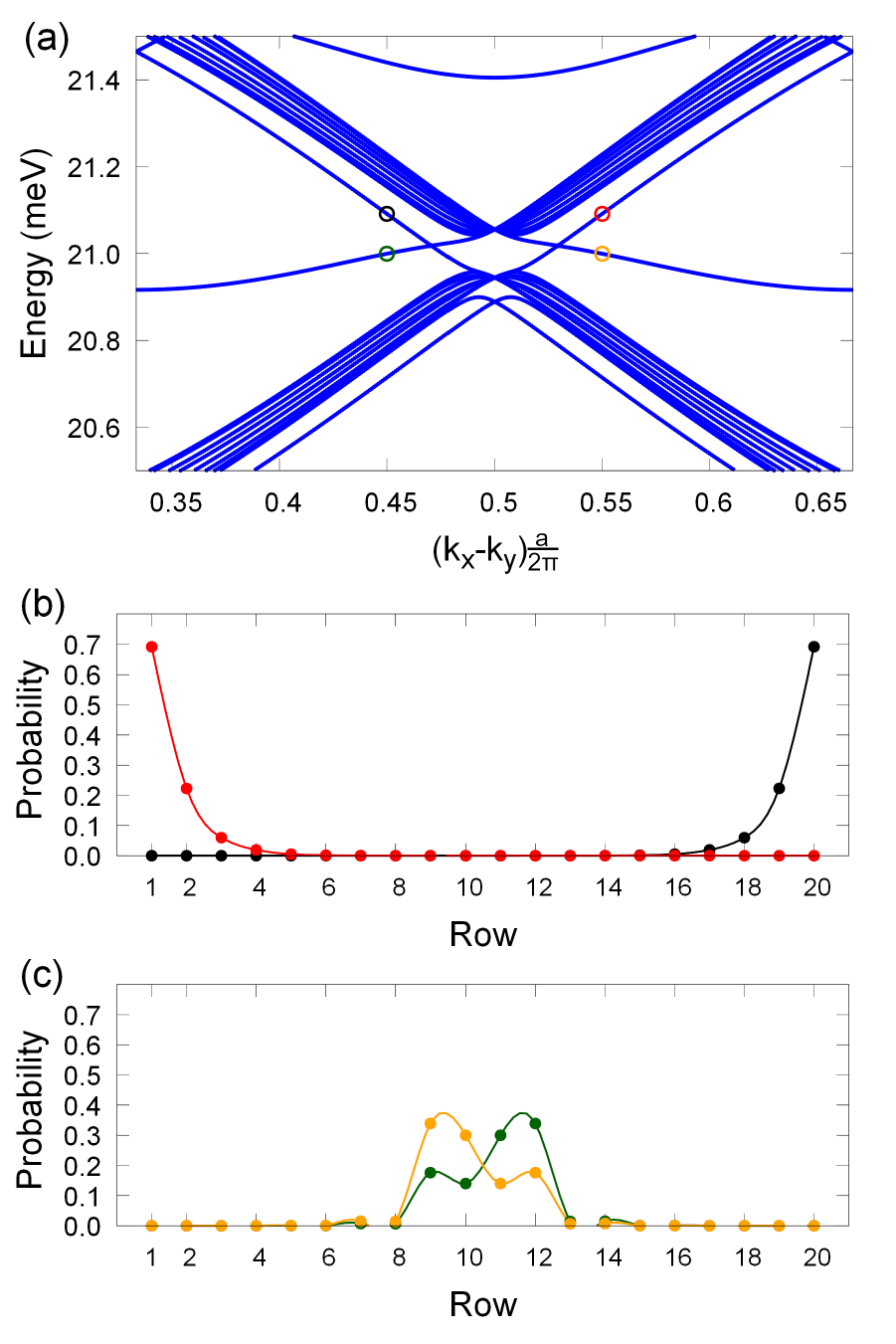} %Fig6.pdf
\end{center}
\caption{(a) Magnon bands for a ribbon of size $M=20$ and $W=3$ and a point defect in the center of the supercell with $J_{1x}=0.5$, $J_{1y}=1.5$, $J_{2x}=8$, and $J_{2y}=8$. (b) Probability distribution for the layer occupation of the edge states marked in (a) in black and red. (c) The same, but for the new state localized near the defect marked in (a) in green and orange.}
\label{Fig6}
\end{figure}

In order to study the effects of defects, we construct a supercell with $W$ columns of $M$ layers each ($W$ is the supercell width measured along the edge of the ribbon). The Hamiltonian matrix $\underline{H}(\vk)$ has dimensions $MW\times MW$. We introduce a defect into a supercell with $M=20$ and $W=3$. The exchange couplings are the same as in Fig.~\ref{Fig5} with the exception that all couplings to the defect site are changed to $40$, making the defect attractive so that it preserves the parallel alignment of the spins.  
This introduces the new mode marked by the green and orange circles in Fig.~\ref{Fig6}(a). As seen from the probability distribution of Fig.~\ref{Fig6}(c), this new mode is tightly connected to the defect.  As expected, this plot indicates that the topological edge mode is not disturbed by the presence of even a strong defect in the center of the ribbon.

\begin{figure}
\begin{center}
\includegraphics[width=7cm]{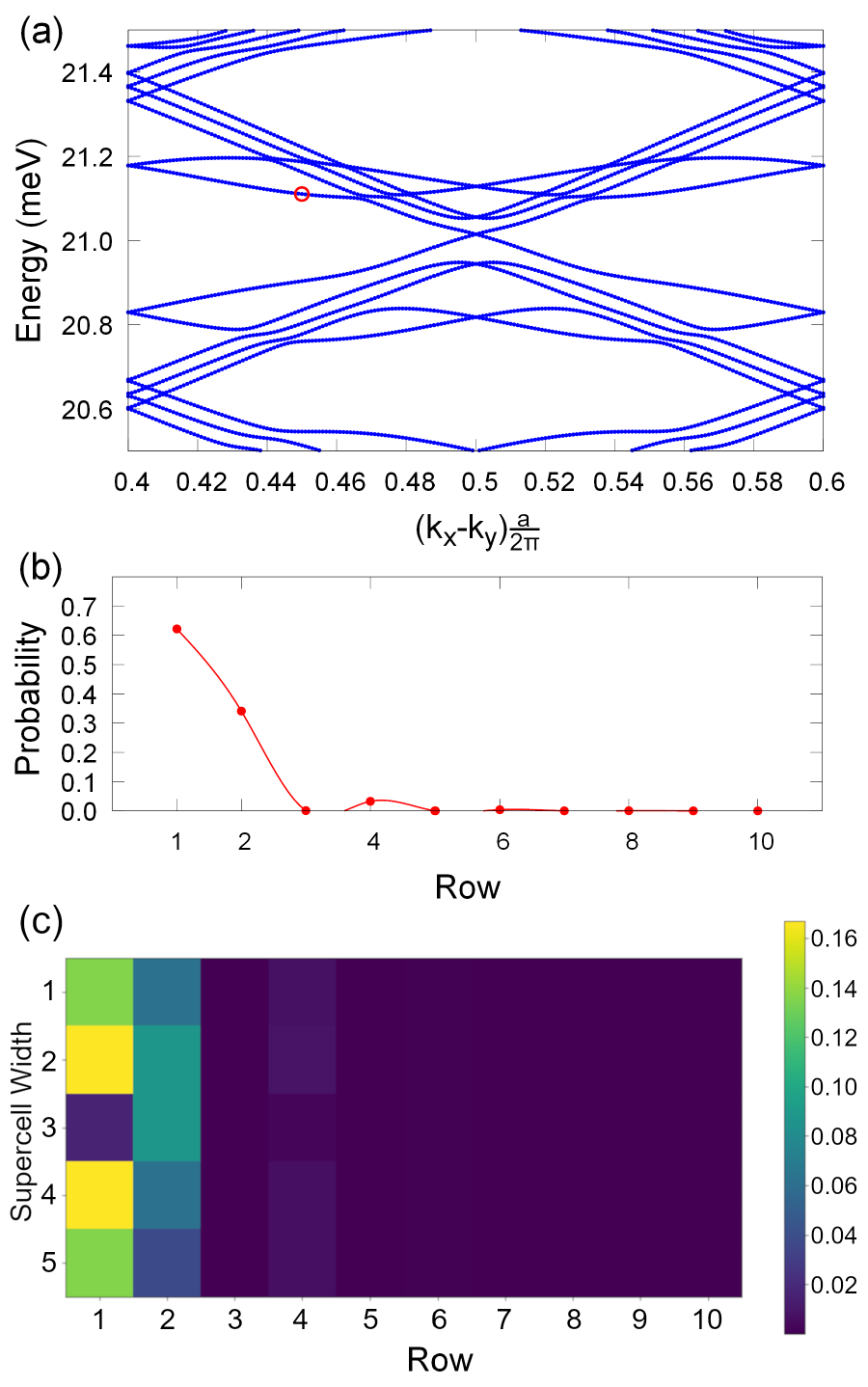}
\end{center}
\caption{(a) Magnon bands for a ribbon of size $M=10$ and $W=5$ and a point defect at the edge with $J_{1x}=0.5$, $J_{1y}=1.5$, $J_{2x}=8$, and $J_{2y}=8$. (b) Probability distribution for the layer occupation of the edge state marked in (a) in red. (c) Probability heat map for the same edge state demonstrating its negotiation of the defect.}
\label{Fig7}
\end{figure}

Additionally we introduce a defect into a supercell with $M=10$ and $W=5$ at one edge. The exchange couplings are again the same as in Fig.~\ref{Fig5} with the exception that all couplings to the defect site at the edge are changed to zero.
This introduces new modes seen in Fig.~\ref{Fig7}(a). As seen from the probability distribution of Fig.~\ref{Fig7}(b), the defect at the edge does have some effect on the topological edge state; there is additional weight at the second atomic layer instead of a rapid and smooth decay into the ribbon. A closer look at the site-specific probability map in Fig.~\ref{Fig7}(c) reveals that the topological edge state still primarily resides at the surface, but it explicitly transports around the edge defect by entering the second atomic layer. This strong local perturbation fails to remove the topological edge state even when imposed at the edge.

\section{Conclusion}
The practical importance of this model is that the topological properties of FM zigzag materials may sensitively depend on strain, which can be used to control the relative anisotropy of the exchange interactions. Suppose, for example, that a pure material has isotropic exchange interactions $J_{1x}=J_{1y}=2$ meV and $J_{2x}=J_{2y}=4$ meV.  Then $\alpha = \beta =1$ and the Chern number is undefined. A small strain modifying the $J_{1}$ exchanges to $J_{1x}=1.9$ meV and $J_{1y}=2.1$ meV while leaving $J_{2}$ exchanges unchanged would lead to $\alpha = 0.905$ and $\beta = 1.033$, causing $\vert C_n\vert =1$ with an edge mode. A small strain modifying the $J_{2}$ exchanges to $J_{2x}=3.9$ meV and $J_{2y}=4.1$ meV while leaving $J_{1}$ exchanges unchanged would lead to $\alpha = 1.105$ and $\beta = 1.033$, causing $\vert C_n\vert =0$ without an edge mode. A small adjustment to the exchange parameters by strain can turn the topologically protected edge modes on/off entirely.

Remarkably, these topological effects are independent of the size of the DM interaction. Even a small DM interaction is enough to open a gap between the magnon bands when the exchange interactions meet the conditions specified in Fig.~\ref{Fig3}. Because strain will usually enhance the symmetry breaking that produces the DM interaction, it is likely to increase the size of $\vert D\vert $ rather than to suppress it.

To summarize, we have identified analytic conditions for the exchange interactions in an anisotropic FM zigzag model to be a Chern topological insulator with Chern numbers of $C_n=\pm 1$.  Because the edge modes of this model are topologically protected from impurities and because the edge modes are sensitive to small changes in the exchange anisotropy, materials that are built on the FM zigzag framework may be important for future technological applications.

\begin{acknowledgments}
We greatly appreciate Dr. Alexander Mook for helpful discussions and constructive criticism during the preparation of the paper.
R.F., L.L., and T.B. were sponsored by the U.S. Department of Energy (DOE), Office of Science, Basic Energy Sciences, Materials Sciences and Engineering Division. S.S. was supported in part by the U.S. DOE, Office of Science, Office of Workforce Development for Teachers and Scientists (WDTS) under the Science Undergraduate Laboratory Internships Program (SULI). J.V. acknowledges startup funding from Middle Tennessee State University.
\end{acknowledgments}

\vspace{3mm}
\begin{centering}
{\bf Data Availability Statement}\\
\end{centering}
The data that support the findings of this study are available from the authors upon reasonable request.

\vfill

\bibliography{mish}

%\vfill\eject

\end{document}